# Near-Field Radiative Heat Transfer between Metamaterials coated with Silicon Carbide Film

Soumyadipta Basu[a)] , Yue Yang, and Liping Wang[b]

School for Engineering of Matter, Transport, and Energy
Arizona State University, Tempe, AZ 85287

## ABSTRACT

In this letter, we study the near-field radiative heat transfer between two metamaterial substrates coated with silicon carbide (SiC) thin films. It is known that metamaterials can enhance the near-field heat transfer over ordinary materials due to excitation of magnetic plasmons associated with *s* polarization, while strong surface phonon polariton exists for SiC. By careful tuning of the optical properties of metamaterial it is possible to excite electrical and magnetic resonance for the metamaterial and surface phonon polaritons for SiC at different spectral regions, resulting in the enhanced heat transfer. The effect of the SiC film thickness at different vacuum gaps is investigated. Results obtained from this study will be beneficial for application of thin film coatings for energy harvesting.

[a]Contact information: soumya.005@gmail.com
[b]Contact information : liping.wang@asu.edu

Near-field radiative heat transfer can exceed blackbody radiation by several orders of magnitude due to the coupling of surface wave excited at the interfaces of two participating media and the vacuum gap separating them[1]. As such, alternative energy systems[2,3], thermal rectifiers[4-6], diodes[7], transistor[8] and a host of other applications have been developed which utilize the near-field enhancement. For nonmagnetic materials the large increase in energy transfer at the near-field is due to the large number of modes excited during surface plasmon resonance for metals and doped semiconductors, as well as due to surface phonon resonance for polar materials both only in case of TM waves ($p$ polarization). In case of metamaterials, surface waves can be excited for both $s$ and $p$ polarization due to negative permeability and permittivity. Several researchers have investigated the energy transfer[9-12], penetration depth[13], local density of states[14-16] as well as thermal rectification[17] during near-field heat radiative transfer between metamaterials.

Previous studies have shown that when the film thickness is on the order of vacuum gap, the near-field heat transfer between them can exceed that between two semi-infinite media due to the coupling of surface polaritons within and between the films[18-21]. Both non-magnetic and metamaterials[12] have demonstrated this phenomenon. However, in all of these cases, the film is either considered to be free-standing or as a thin film of a nonmagnetic material on a nonmagnetic substrate. In this study we investigate the near-field heat transfer between two metamaterials which are coated with SiC thin films. The effect of film thickness at different vacuum gaps is investigated and the total energy transfer between these structures is compared to that between two bulk metamaterials. Through detailed analysis we show that depending on the vacuum gap and thickness, the SiC film can be transparent to the thermal emission from the metamaterial which helps to couple the magnetic and electric surface plasmon polaritons (SPPs) between the metamaterials. In addition, the surface waves excited at SiC-vacuum interface can also couple thereby increasing the overall heat transfer.

Figure 1 shows the schematic for the near-field radiative heat transfer between two semi-infinite metamaterials coated with SiC thin films and separated by a vacuum gap, $d$. The emitter is at temperature $T_1=300$ K and receiver is at temperature $T_2 =0$ K. The film thicknesses are denoted by $t_1$ and $t_3$ respectively. The semi-infinite emitter is represented as media 0, the thin film on the emitter side as media 1, vacuum gap as media 2, thin film on the receiver side as media 3, and the semi-infinite receiver as 4. Therefore we are considering a 5-layer structure in this study. Optical properties of SiC were taken from Palik's data[22] while the metamaterial is considered to be made of thin wires and split ring resonators as used in Refs. 9 and 13. Note that the metamterial chosen in this study is just an illustrative example and the analysis performed here is structure invariant. The permittivity and permeability models for the metamaterial are obtained from Ref. [13] as shown below.

$$\varepsilon_r = (\varepsilon'_r + i\varepsilon''_r) = 1 - \frac{\omega_p^2}{\omega^2 + i\omega\gamma_e} \tag{1}$$

$$\mu_r = (\mu'_r + i\mu''_r) = 1 - \frac{F\omega^2}{\omega^2 - \omega_0^2 + i\omega\gamma_m} \tag{2}$$

where $F = 0.56$ is the split ring filling factor, $\omega_p = 10^{14}$ rad/s is an equivalent plasma frequency, $\omega_0 = 0.4\omega_p$, and $\gamma_e = \gamma_m = \gamma = 0.01\omega_p$.

The expression for the radiative heat flux between any two media is[1]

$$q'' = \frac{1}{\pi^2} \int_0^\infty d\omega \left[\Theta(\omega,T_1) - \Theta(\omega,T_2)\right] \int_0^\infty s(\omega,\beta) d\beta \tag{3}$$

Here, $\beta$ is the horizontal component of the wavevector, $\Theta(\omega,T) = \hbar\omega/[\exp(\hbar\omega/k_B T)-1]$ is the mean energy of a Planck oscillator at thermal equilibrium temperature $T$, where $\hbar$ and $k_B$ are the reduced Planck constant and the Boltzmann constant, respectively. In the near-field, propagating waves ($\beta < \omega/c$) have negligible contribution to the heat transfer, hence only contributions from evanescent waves ($\beta > \omega/c$) are considered. The exchange function $s(\omega,\beta)$ is then expressed as[23]

$$s_{evan}(\omega,\beta) = \frac{\text{Im}(R_{210}^s)\text{Im}(R_{234}^s)\beta e^{-2\text{Im}(\gamma_2)d}}{\left|1 - R_{210}^s R_{234}^s e^{i2\gamma_2 d}\right|^2} + \frac{\text{Im}(R_{210}^p)\text{Im}(R_{234}^p)\beta e^{-2\text{Im}(\gamma_2)d}}{\left|1 - R_{210}^p R_{234}^p e^{i2\gamma_2 d}\right|^2} \quad (4)$$

where $\gamma$ is the vertical-component wavevector, $R_{210(234)}$ is respectively the reflection coefficient from the vacuum gap to either emitter (i.e., $R_{210}$) or to the receiver (i.e., $R_{234}$), and can be calculated by thin-film optics for both $s$ and $p$ polarizations[24,25]

$$R_{234(210)} = \frac{r_{23(21)} + r_{34(10)} e^{i2\gamma_{3(1)} t_{3(1)}}}{1 + r_{23(21)} r_{34(10)} e^{i2\gamma_{3(1)} t_{3(1)}}} \quad (5)$$

Note that when the thin films (i.e., layer 1 or 3) become semi-infinite, the thin-film reflection coefficient $R_{210}$ (or $R_{234}$) will be replaced by the reflection coefficient $r_{21}$ ($r_{23}$) at a single interface. In addition, if the film substrate (i.e., layer 0 or 4) is vacuum, the energy associated with propagating waves will not be absorbed when it transmits into the vacuum substrate.

Figure 2 is a contour plot of the $s$ function at two different film thicknesses when the vacuum gap between the receiver and emitter is set to 10 nm. The contour is plotted as a function of angular frequency ($\omega$) and $\beta^*$ ($\beta c/\omega$). Dispersion relation for the 5-layer media is also plotted in Figs. 2(a) and (b) as shown by the dotted lines.

Next we outline the method of deriving the dispersion relation for TM waves for the structure under consideration. Dispersion relation for TE waves can be derived using the same procedure by using the appropriate Fresnel reflection coefficients.

For TM waves, the SPP dispersion relation can be obtained by the following:

$$1 - R^p_{210} R^p_{234} e^{i2\gamma_2 d} = 0 \tag{6}$$

For evanescent waves when $\beta \gg \omega/c$,

$$R^p_{234(210)} = \frac{r^p_{23(21)} + r^p_{34(10)} e^{-2\beta t_{3(1)}}}{1 + r^p_{23(21)} + r^p_{34(10)} e^{-2\beta t_{3(1)}}} \tag{7}$$

Substituting values of the reflection coefficients in Eq. (6) and after simplification we get

$$\left[ (\varepsilon_{SiC}+1)(\varepsilon_{SiC}-\varepsilon_m) - (\varepsilon_{SiC}-1)(\varepsilon_{SiC}+\varepsilon_m) e^{-2\beta t_1} \right] \left[ (\varepsilon_{SiC}+1)(\varepsilon_{SiC}-\varepsilon_m) - (\varepsilon_{SiC}-1)(\varepsilon_{SiC}+\varepsilon_m) e^{-2\beta t_3} \right] e^{-2\beta d}$$
$$= \left[ (\varepsilon_{SiC}+1)(\varepsilon_{SiC}+\varepsilon_m) - (\varepsilon_{SiC}-1)(\varepsilon_{SiC}-\varepsilon_m) e^{-2\beta t_1} \right] \left[ (\varepsilon_{SiC}+1)(\varepsilon_{SiC}+\varepsilon_m) - (\varepsilon_{SiC}-1)(\varepsilon_{SiC}-\varepsilon_m) e^{-2\beta t_3} \right] \tag{8}$$

where $\varepsilon_{SiC}$ and $\varepsilon_m$ are the dielectric functions for SiC and metamaterial respectively. Eq. (8) can be further rewritten as

$$h_1 x^2 y^2 + h_2 x^2 + h_3 y^2 + h_4 xy + h_5 = 0 \tag{9}$$

$h_1 = e^{\beta(d-t_1-t_3)}$, $h_2 = e^{\beta(-d-t_1+t_3)}$, $h_3 = e^{\beta(-d+t_1-t_3)}$, $h_4 = -[e^{\beta(d-t_1+t_3)} + e^{\beta(d+t_1-t_3)} + e^{\beta(-d+t_1+t_3)} + e^{\beta(-d-t_1-t_3)}]$,

$h_5 = e^{\beta(d+t_1+t_3)}$, $x = (\varepsilon_{SiC}-1)/(\varepsilon_{SiC}+1)$, and $y = (\varepsilon_{SiC}-\varepsilon_m)/(\varepsilon_{SiC}+\varepsilon_m)$.

Equation 9 is then solved numerically to obtain the dispersion relation as plotted in Figs. 2(a) and (b). For simplicity in this study we consider film with same thicknesses. In other words $t_1=t_3=t_f$. Peaks predicted by the dispersion relation matches exactly with those from the contour plots. Notice that the two peaks at around 3 and $4.5 \times 10^{13}$ rad/s correspond to the electric and magnetic resonance for emission from the metamaterial substrate while the strong peak at around $1.8 \times 10^{14}$ rad/s is due to the coupling of evanescent waves emitted from the SiC film. For 10 nm SiC film, we can clearly see the splitting of the near-field flux into the symmetric and anti-symmetric resonance modes due to the coupling of the surface waves generated at the vacuum – film interface within and between the two SiC films. Notice that the spectral locations of the electric surface plasmons (ESP) and magnetic surface plasmons (MSP) for the emission from the metamaterial when it is coated with SiC doesn't match with those without any coating as seen in Ref.[13]. This is because the reflection coefficient in Eq. (4) for both $s$ and $p$ polarization involves contribution from the SiC thin film as well. When the SiC film thickness increases from 10 to 100 nm, the strength of the resonance due to metamaterial decreases while that from the SiC increases. This is because at 10 nm vacuum gap, the 100 nm thick SiC film behaves as a semi-infinite media and nearly absorbs all the emitted radiation from the metamaterial. Figure 2 therefore shows that it is possible to tune the strength and location of electric and magnetic resonance from a metamaterial by just adding a 10 nm thin film on the substrate. In addition, coupling of the surface waves between and within the SiC film further increases the channels of heat transfer which can increase the heat transfer over that between two metamaterials.

Figure 3 plots the ratio of the total near-field heat transfer between SiC coated metamaterials ($Q''$) to bulk metamaterial ($Q''_{bulk}$) for different film thicknesses as a function of the vacuum gap. For comparison the total heat transfer between two semi-infinite metamaterials is also plotted in Fig. 3. Interestingly, the impact of the film thickness on the overall heat transfer grows stronger as the film thickness decreases with $Q''$ nearly four times that between two bulk metamaterials. For $t_f \leq d$, the SiC film is transparent to the thermal emission from the metamaterial substrate as the penetration depth of thermal radiation is greater than film thickness[26]. In addition the splitting of the near-field flux into 4 different resonant modes provides additional channels of heat transfer. On the other hand, for $t_f \geq d$, the thermal emission from the substrate is mostly absorbed inside the film due to the small penetration depth which results in the reduction in the total near-field heat transfer as compared to bulk metamaterials. At the same time instead of 4 there are only 2 resonant modes which further reduce the overall near-field heat transfer[18]. Therefore it is possible to tune the near-field heat transfer between the two metamaterials by using a thin film. Figure 3 further shows that for 1 nm film, $Q''/Q''_{meta}$ becomes saturated for $d > 20$nm. For 10 nm similar phenomena can be seen at larger vacuum gaps. For $t_f = 1$nm, $Q''$ increases as the vacuum increases as the inter- and intra-film coupling of surface waves becomes stronger as explained earlier. Beyond 20 nm, the coupling becomes constant and the rate of change of $Q''$ with vacuum gap is same as that of $Q''_{meta}$ resulting in $Q''/Q''_{meta}$ becoming independent of vacuum gap for $d > 20$nm. Figure 3 clearly shows the advantage of adding even a 1 nm thin SiC film on the metamaterial to further enhance the near-field heat transfer.

The contribution of the SiC film and the underlying metamaterial substrate to the near-field heat transfer can be further understood from the spectral plot of the near-field heat flux between the emitter and the receiver as shown in Figs. 4(a) and (b) for 10 and 100 nm vacuum gaps respectively. Again the spectral heat flux between the two semi-infinite metamaterials is plotted for comparison. As explained earlier, the SiC film shifts the spectral location for the electric and magnetic resonance for emission from the metamaterial. Having the SiC thin film on top of the metamaterial enables an additional spectral peak corresponding to the surface phonon polariton coupling between the SiC films. Based on the vacuum gap and film thickness some interesting observations can be made from Figs 4(a) and (b) as discussed next. For $t_f/d = 0.01$ as in Fig. 4(b) the spectral peak due to MSP from the metamaterial matches exactly with the one from bulk metamaterial. However the spectral location of ESP is shifted to shorter frequency as compared to the bulk. When $t_f/d = 0.1$ as in Figs 4(a) and (b), the m-SP and e-SP for metamaterials merge into a single peak. For $t_f/d > 0.1$, the ESP and MSP due to the metamaterial again show up, however, the frequencies at which the SPs are excited in the metamaterial are closer than when the metamaterial doesn't have the SiC film. This is again due to the effect of the SiC film on the reflection coefficients. In summary, the impact on thermal emission from the metamaterial becomes more profound as the SiC film thickness increases. For SiC film as is well known, when $t_f/d <1$, there are two spectral peaks for SiC, corresponding to the symmetric and anti-symmetric mode with the former being more dominant as discussed in previous studies[18]. As the film thickness increases and is greater than the vacuum gap, the two modes merge into a single spectral peak.

In conclusion, we have shown that the near-field heat transfer can be enhanced over that between two bulk metamaterials by adding a thin film of SiC on top of both semi-infinite metamaterial emitter and receiver. By taking advantage of excitation of surface plasmon resonance in metamaterial substrate and surface phonon polaritons in SiC film at different frequencies it is possible to exceed the total heat transfer over that between bulk metamaterials. This can be achieved when the film thickness is on the order of vacuum gap. Results obtained from this study will facilitate experiments on measurement of near-field heat flux between two metamaterials.

**Figure Captions:**

Fig. 1    Schematic of the near-field radiative heat transfer between two metamaterials coated with SiC. The emitter including the metamaterial and the SiC film (media 0 and 1) is at 300 K and the receiver is at 0 K. The receiver and the emitter are separated by a vacuum gap of thickness $d$.

Fig. 2    Exchange function $s(\omega,\beta)$ for the proposed structure as a function of $\omega$ and $\beta$ for (a) 10 nm film and (b) 100 nm SiC film thickness. In both cases, the vacuum gap is set to 10 nm. The dashed lines in the two figures represent the dispersion curves.

Fig. 3    Total radiative heat flux between the SiC coated metamaterials as a function of vacuum gap for different film thicknesses. For comparison, the heat flux between two metmaterials without the SiC coating is also plotted.

Fig. 4    Spectral heat flux between SiC coated metamaterials for (a) 10 nm and (b) 100 nm vacuum gap for different film thicknesses.

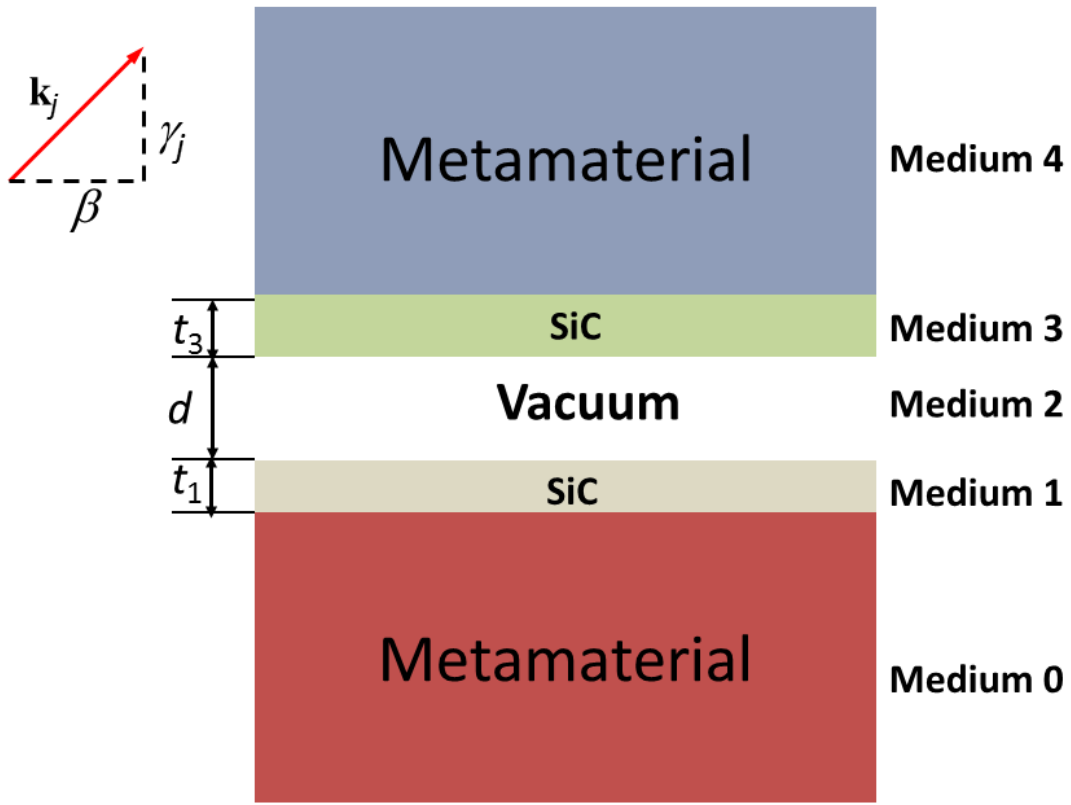

Figure 1

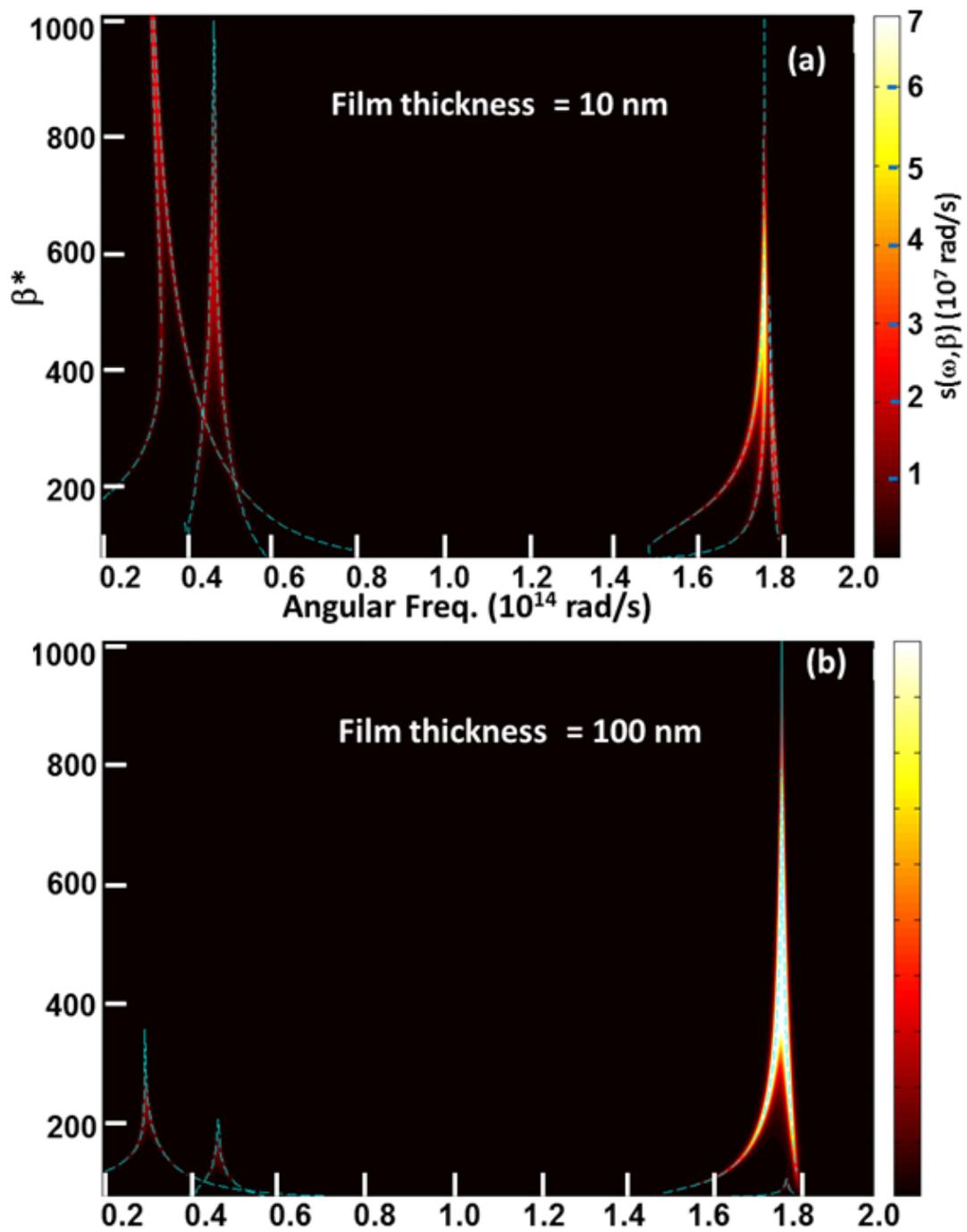

Figure 2

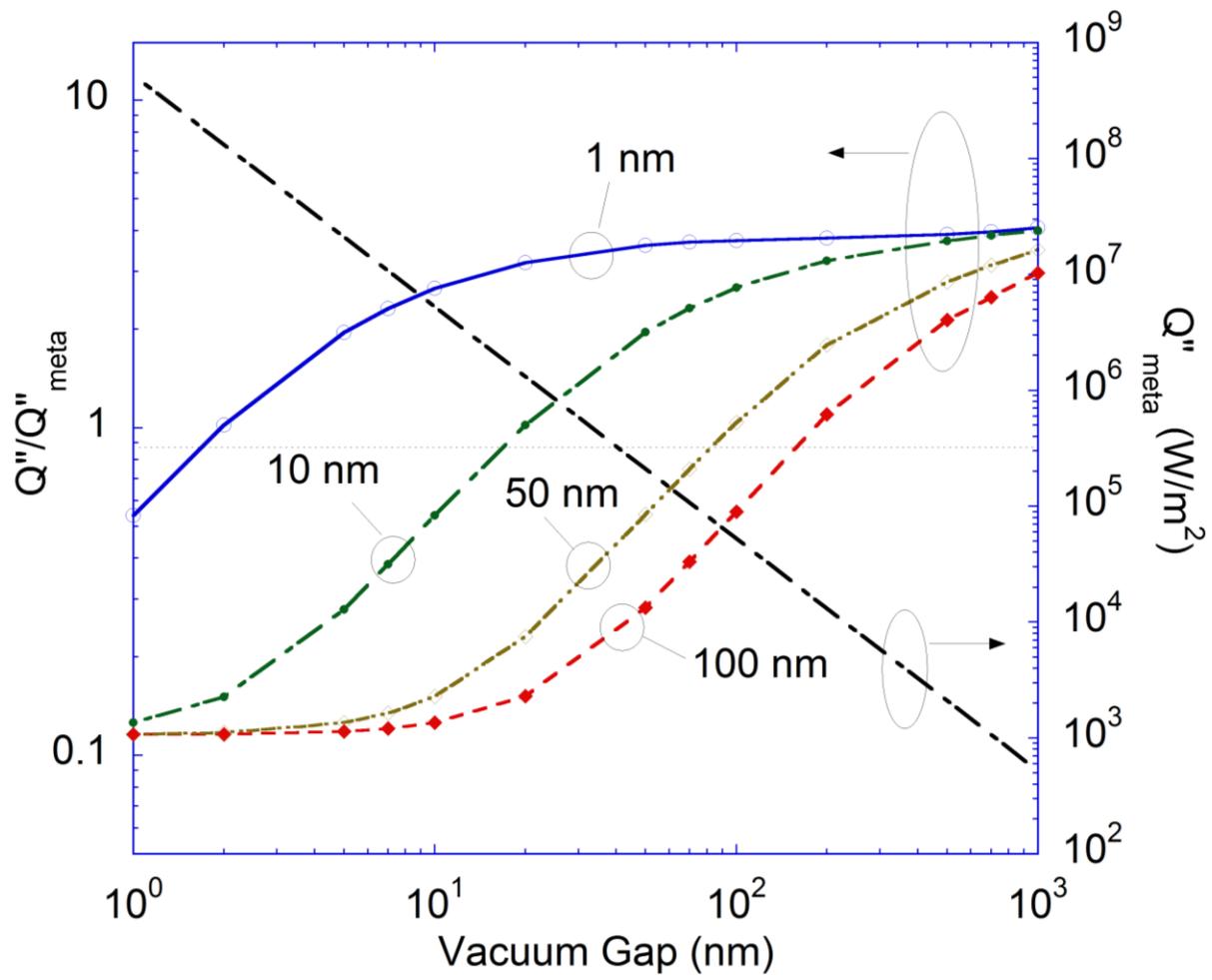

Figure 3

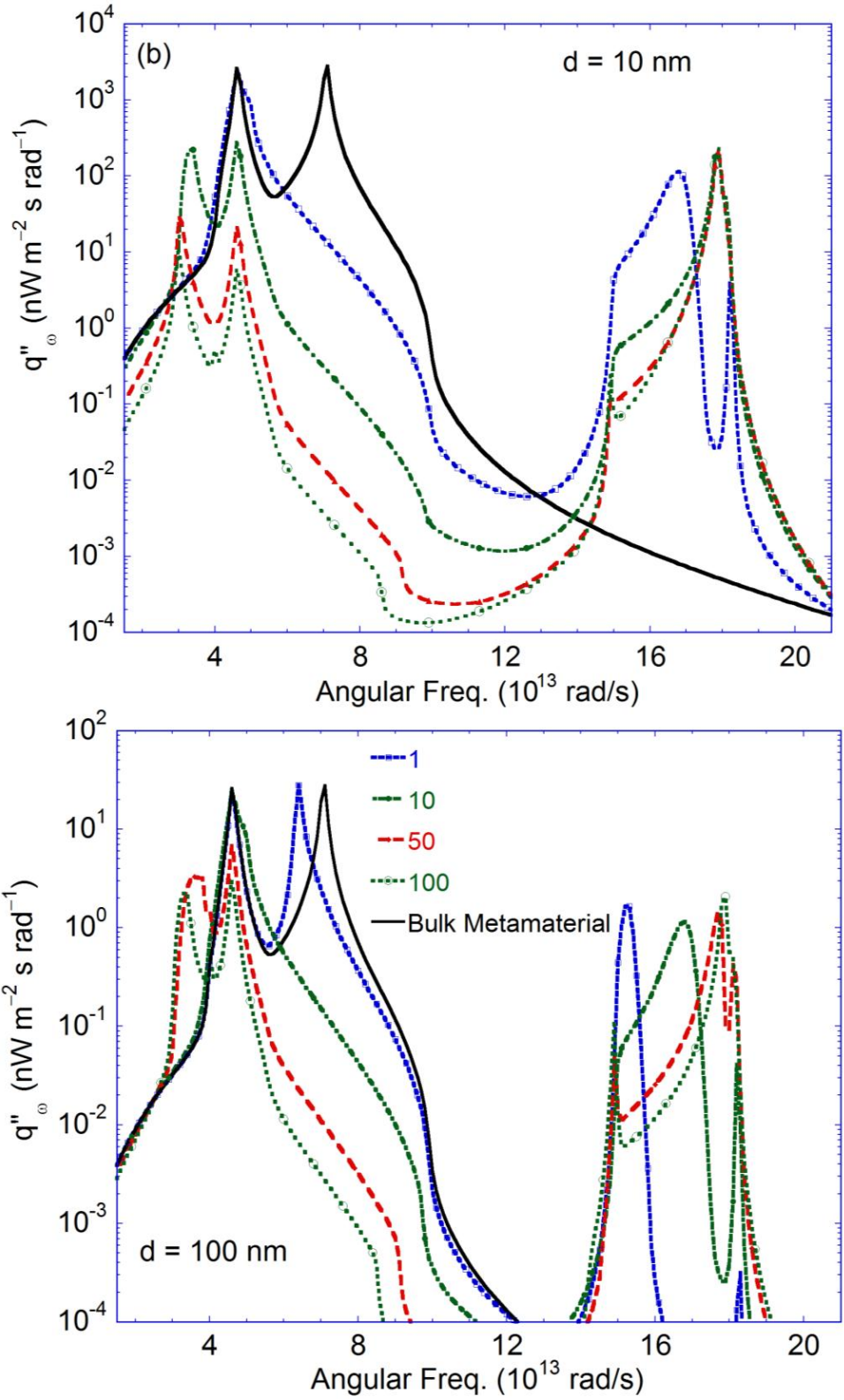

Figure 4